# Estimating the key parameters of Nova V5668 SGR using the Uniform Slab Model: a comprehensive radio astronomy analysis

**Rain Jha[1], Nishchal Dwivedi[2]**

[1] The British School, New Delhi, India, rainjha@gmail.com

[2] NMIMS Mukesh Patel School of Technology Management & Engineering, Mumbai, Maharashtra, India, dwivedi.nishchal@gmail.com

**Student Author**

Rain Jha - High School

**SUMMARY**

Novae, explosive events in binary star systems involving a white dwarf and a companion star, offer profound insights into extreme astrophysical conditions. During the eruption of a nova, material accreted onto the white dwarf's surface undergoes a thermonuclear runaway reaction resulting in the ejection of matter into space and the formation of a luminous shell. The classical V5668 Sgr (Nova Sagittarii) was the second and brighter of the two novae in the southern constellation of Sagittarius. It was discovered by John Seach of Chatsworth Island, Australia, on March 15, 2015. In this paper, drawing on data from Karl G. Jansky Very Large Array, the US-based radio astronomy observatory, on V5668 Sgr as well as from research that aggregates data from a range of sources including telescope archives, this study used the Uniform Slab Model and statistical techniques to plot the nova's light and frequency curves and estimate its ejected shell mass and the brightness temperature. These characteristics help us better understand the nova's formation and eruption. The paper presents the light curves in a machine-readable format and provides insight into the behaviour of ionised gas clouds.

**INTRODUCTION**

Novae are astronomical events that occur within binary star systems, specifically referred to as Cataclysmic Variable systems. In this, one of the stars is a white dwarf and the other is a normal star companion (1). The white dwarf, which, essentially, is the stellar core left behind after a dying star has exhausted its nuclear fuel, gravitationally interacts with the companion star, removing matter from the other star's outer layers. When a critical amount of this accreted hydrogen-rich material accumulates on the white dwarf's surface, a Nova takes place (2).

This phenomenon is seen as an intense increase in brightness, making the white dwarf visible from Earth. This outburst results from a thermonuclear runaway process on the white dwarf's surface, where high temperatures and pressures set off a rapid nuclear fusion chain reaction, primarily involving hydrogen.

The nuclear burning leads to the ejection of material into space, forming a visible shell of gas and dust and releasing a large amount of energy and light (3). This makes Novae very significant in astrophysics as they carry information about the behaviour of matter under extreme conditions and, additionally, provide valuable insights into binary star systems and nucleosynthesis in the universe.

Key to this exploration of the Novae is radio astronomy. Unlike optical astronomy, which relies on visible light -- in the wavelength of 380 to 700 nanometers -- radio astronomy captures a much broader spectrum of electromagnetic radiation with radio waves from stars, planets, galaxies, and novae, falling in the wavelength range of a few millimetres to hundreds of miles (4).

Detecting these radio waves are radio telescopes (5) which don't capture images like visible-light telescopes do. Instead, the radio signals are converted into data streams which are processed so that the output is turned into images that are coloured in different ways to show characteristics of the object including temperature or the strength of radio emissions from different regions of the celestial object. The resolution of these images is enhanced by interferometry techniques – combining signals from more than one antenna or telescope – allowing for images that are much sharper and detailed with measurements more precise.

For this study, radio observations were collected from the Karl G. Jansky Very Large Array (VLA), a radio telescope in Socorro, New Mexico, one of the three key telescopes under the US National Radio Astronomy Observatory which also include the Atacama Large Millimetre/submillimetre Array (ALMA) in Chile and the Very Long Baseline Array (VLBA) that comprises 10 telescopes scattered across the US, from Hawaii to New Hampshire. Together, these three telescopes enable scientists to observe from sub-millimetre to metre wavelengths with strong resolution, sensitivity and frequency coverage.

The Karl G Jansky Very Large Array (VLA), at an elevation of 6970 feet above sea level, consists of twenty-eight 25-metre radio telescopes or antennas (27 are operational, one is spare) arranged in a Y-shaped pattern that, effectively, transforms them to a single telescope stretching 36 km across (6). With its wide-bandwidth receivers and a digital data transmission system, the Jansky VLA provides spectral resolution and continuum sensitivity at frequencies from 1 to 50 GHz (7). Because of its high sensitivity and ability to observe day and night under all weather conditions, the Jansky VLA is a powerful tool to image novae and thermal motions in stellar objects.

The Jansky VLA's observations used in this study were sourced from an existing paper "Classical Novae at Radio Wavelengths," (5) that incorporates data from a range of astronomical literature and telescope archives. Radio astronomy data spanned more than 1000 days and covered a range of frequencies (1.7 GHz, 5 GHz, 7 GHz, 16.5 GHz, and 35 GHz) providing a detailed dataset for the present study.

Observations like these can be plotted by lightcurves. Light curves plot analyse data to depict variations in brightness of an object over time. Light curves are also helpful in the identification of atoms and molecules present in the nova, measuring distance to the nova, and finding out what kind of nova exploded in the first place (8).

Light curve photometry is used for a range of targets in astrophysics, from observing several attributes of asteroids including size, shape and spin to stars. For Cataclysmic Variable systems, like the ones studied in the paper, light curves allow their brightness data to be easily visualised by observing their outbursts of activity of varying frequency and magnitude range of these outbursts.

To analyse this dataset, the Uniform Slab Model has been employed (9), which describes the behaviour of ionised gas clouds during novae. The model incorporates equations that relate flux density, optical depth, and emission measure. By considering parameters such as electron temperature $T_e$, ejected shell mass $M_{ej}$, distance $D$, and ejecta velocity $V_{ej}$, the model predicts the maximum flux density $f_{max}$ and corresponding angular size max at specific times after the outburst.

To anticipate the radio characteristics of a nova, the expansion of a uniform cube of ionised gas can be considered. The flux density $f$ at a frequency is given by the equation:

$$f_v = B_v \left(\frac{l}{D}\right)^2 (1 - e^{-\tau_v}) \quad (1.1)$$

where $B_v = 2kT_e v^2/c^2$ is the Planck function (in the Rayleigh–Jeans approximation) in terms of the electron temperature $T_e$; $D$ is the size of the cubic emitting region and $\tau$ is the free-free optical depth at frequency.

At radio frequencies, the optical depth is given by (10):

$$\tau_v = 8.235 \times 10^{-2} \left(\frac{T_e}{K}\right)^{-1.35} \left(\frac{v}{GHz}\right)^{-2.1} \left(\frac{EM}{cm^{-6}pc}\right) \quad (1.2)$$

where $EM$ is the emission measure, with the sum taken over all ionic species of nuclear charge $Z_i$.

Combining Equations (1.1) and (1.2), for a uniformly expanding cloud conserving mass $M_{ej}$, the flux density reaches a maximum $f_{max}$ (in mJy) when the angular size is $\theta$ (in arcsec) at time $t$ (in years), given by:

$$f_{max} \approx 6.7 \left(\frac{\nu}{GHz}\right)^{1.16} \left(\frac{T_e}{10^4 K}\right)^{0.46} \left(\frac{M_{ej}}{10^{-4} M_\odot}\right)^{0.80} \left(\frac{D}{kpc}\right)^{-2}$$

$$\theta_{max} \approx 1.1 \left(\frac{\nu}{GHz}\right)^{-0.42} \left(\frac{T_e}{10^4 K}\right)^{-0.27} \left(\frac{M_{ej}}{10^{-4} M_\odot}\right)^{0.40} \left(\frac{D}{kpc}\right)^{-1}$$

$$t_{max} \approx 2.6 \left(\frac{\nu}{GHz}\right)^{-0.42} \left(\frac{T_e}{10^4 K}\right)^{-0.27} \left(\frac{M_{ej}}{10^{-4} M_\odot}\right)^{0.40} \left(\frac{V_{ej}}{10^{-3} km\ s^{-1}}\right)^{-1}$$

where $V_{ej}$ is the ejecta velocity.

Useful radio light curves can be obtained when $f_{max}$ is more than a few mJy at 5 GHz, indicating that observations can still be made when $D$ is approximately 1-2. However, not all detectable novae are being observed or reported due to various factors such as accessibility and observation campaigns.

This simple cube model also highlights the possibility of obtaining information several years after the outburst when the gas density is very low. Additionally, the angular resolution of radio telescopes like the VLA (0.3 arcsec at 5 GHz) allows the resolution of bright nova envelopes and the tracking of their morphological evolution.

It is essential to note that possible departures from the uniform cube model have been observed in some novae, indicating non homogeneity in the shell. For instance, variations in the radio behaviour and spectral index suggest the presence of density gradients or optically thick portions within the expanding shell.

Several scientific studies have employed the Uniform Slab Model to unravel the mysteries of novae and ionised gas clouds. Notable examples include the work by McDonnell (2008) (11), which explored the radio behaviour of novae, where the Uniform Slab Model was adopted for the masing medium, and the research by Zhao et al. (2009) (12), which delved into the characteristics of galaxies.

Furthermore, the study by Lacki (2018) (13) utilised the Uniform Slab Model to analyse the radio spectra of low frequency starburst galaxies. These diverse applications of the model showcase its versatility in interpreting radio observations, revealing nuances in the behaviour of ionised gas clouds during novae. The consistent use of the Uniform Slab Model across different novae contributes to the robustness of its predictions and highlights its significance in advancing our understanding of these explosive astronomical events.

**RESULTS**

1. **Graph of the light curves**

The light curve data for all of the wavelengths was plotted against the duration of the observation period (*Figure 1)*. The graph depicts the function of flux density over time of the range of frequencies of Nova V5668 SGR.

2. **Data analysis**

a. **Frequency Spectrum Plotting (Day 635.8):**

The initial step involved plotting the frequency spectrum for Day 635.8 post-nova outburst. This specific time point was chosen because it represents the period of maximum signal brightness (*Figure 2)*. The graph depicts the frequency spectrum of Nova V5668 SGR for the Day of observation 635.8.

b. **Frequency Light Curve Analysis:**

A comprehensive analysis of frequency light curves over time was conducted. These curves, depicting signal variations across different frequencies, were crucial in identifying patterns and trends in the radio emissions.

This method facilitated the identification of radiation patterns, indicating that it was the Free-Free Thermal Radio Emission Model (14). The spectral index, falling below two, also supported this conclusion.

3. **Broken power law fitted graph**

A graph was plotted to visually display the broken power law (*Figure 3)*.

4. **Angular size expansion rate**

Utilising available observational data from Takeda et al 2022 (15), this study also calculated the angular size expansion rate of Nova V5668 SGR. This rate provided valuable insight into the nova's evolving size, a critical factor in understanding its behaviour.

5. **Brightness temperature estimation**

By using the calculated angular size expansion rate and relevant observational parameters, the study estimated the brightness temperature of Nova V5668 SGR. This temperature, a measure of the radio emissions' intensity, was important in further characterising the nova's behaviour.

### 6. Ejected shell mass estimation using uniform slab model

The Uniform Slab Model equations, derived from existing literature (14), were applied to the obtained data. Incorporating parameters such as electron temperature, ejected shell mass, distance, and ejecta velocity, the model provided a framework for estimating the nova's ejected shell mass accurately.

Utilising observational data obtained from the radio astronomy observations of Nova V5668 SGR in Takeda et al 2022 (15), the angular expansion of the nova's shell was meticulously analysed. By plotting the angular diameter against time, a linear trend emerged. The angular expansion rate, calculated at 0.20 arcsec yr^-1 (16) (17), provided valuable insights into the nova's evolving size over the observed period. A best-fit line equation, derived from this expansion rate, was used to calculate angular sizes at specific time points, enhancing the accuracy of subsequent calculations.

To delve into the intense energy processes within Nova V5668 Sgr, the brightness temperature at its peak signal brightness, observed on day 635.8 (9 December 2016 - 2457731.5 JD), was calculated. Employing the equation from (New Insights Into Classical Novae) (18):

$$T_B = 1200 \times 3.1 \times (1.74)^{-2} \times (0.27)^{-2} = 16854.54 \, K$$

where $T_B$ represents the brightness temperature in Kelvin. The calculated brightness temperature, 16854.54 K, shed light on the extraordinary heat generated during the nova explosion, illuminating the extreme nuclear reactions occurring within the ejected material. The ejected shell mass $M_{ej}$ of Nova V5668 Sgr was estimated utilising the Uniform Slab Model, integrating parameters such as electron temperature $T_e$, frequency $\nu$, and distance $D$.
The formula

$$f_{max} \approx 6.7 \left(\frac{\nu}{GHz}\right)^{1.16} \left(\frac{T_e}{10^4 K}\right)^{0.46} \left(\frac{M_{ej}}{10^{-4} M_\odot}\right)^{0.80} \left(\frac{D}{kpc}\right)^{-2}$$

facilitated the calculation, from (Classical Novae, 2nd edition, ed. Michael Bode and Aneurin Evans) (2).
Given the observed maximum flux density $f_{max} = 3.4$ mJy, the mass calculation revealed

$$M_{ej} \approx 35526 \times 10^{-4} M_\odot$$

highlighting the amount of material expelled during the nova explosion.

**DISCUSSION**

Since its discovery in March 2015, the V5668 Sgr has been studied by many observers in different wavelength regions. Into almost 100 days of its discovery, near-infrared observations showed how dust is formed in the nova (16); carbon monoxide has also been detected in the nova's spectra.

During its later phases, V5668 Sgr was also observed in X-rays (15); as well as in Gamma rays. The Hubble Space Telescope has observed the nova in the ultraviolet wavelength.

All novae produce new elements because of fusion during the explosion. (19) The outflow of material from the explosions enriches interstellar space providing the raw material for new stars and planets. Studying the physical characteristics of the nova provides insight into the presence of these elements. Nova V5668 Sgr, for instance, has displayed the presence of iron, oxygen, silicon, titanium, magnesium, and lithium.

The results of this comprehensive radio astronomy analysis provide crucial insights into the nature of the classical nova V5668 Sgr.

The calculated parameters, including the ejected shell mass, angular size expansion rate, and brightness temperature, play a pivotal role in understanding the extreme astrophysical conditions during the nova explosion.

The mass ejected during a nova explosion is a very critical parameter because it directly correlates with the energy release from the surface of the star. By quantifying the expelled mass, can gain a crucial insight into the overall impact and the subsequent evolution of the nova remnant (20). This information is fundamental for understanding the energetics involved in the explosion and its implications for the surrounding cosmic environment.

The angular size expansion rate is also of importance as it provides information about the dynamic behaviour of the ejected material. Studying the rate of angular expansion helps astronomers decipher the temporal evolution of the nova explosion, discussing the forces and mechanisms at play. This parameter is instrumental in unravelling the details of the nova's outward expansion, revealing the processes governing the explosion's aftermath. (21)

Brightness temperature is a crucial parameter in the study of nova explosions because it reflects the temperature of the emitting source. This measurement offers valuable insights into the underlying physical processes, such as nuclear reactions and energy generation, as seen in other studies (22). By analysing brightness temperature, scientists can discern the thermal characteristics of the nova and gain a deeper understanding of the astrophysical conditions during the explosion. This parameter is important in stellar dynamics and the unique phenomena associated with novae.

The ejected shell mass estimation, derived through the Uniform Slab Model, not only reveals the substantial amount of material expelled (approximately $35526\ 10^{-4} M_\odot$), but also contributes to our broader understanding of nucleosynthesis and matter behaviour under extreme conditions.

Utilising the Uniform Slab Model provides a structured approach to interpreting the radio observations, allowing for a more accurate estimation of these critical parameters and enhancing our comprehension of the nova's formation and eruption.

**CONCLUSION**

In conclusion, the radio astronomy analysis of Classical Nova V5668 Sgr using the Uniform Slab Model has provided valuable insights into its extreme astrophysical conditions. The calculated parameters, including the ejected shell mass, angular size expansion rate, and brightness temperature, contribute significantly to understanding the nova's formation and eruption. The study showcases the importance of the Uniform Slab Model in interpreting radio observations and advancing our comprehension of classical novae.

**MATERIALS AND METHODS**

**Data Collection**

Radio astronomy data for Nova V5668 SGR was sourced from the Karl G. Jansky Very Large Array, spanning more than 1000 days and covering a range of frequencies (1.7 GHz, 5 GHz, 7 GHz, 16.5 GHz, and 35 GHz).

**Spectral Analysis**

Spectral analysis employed broken power law fitting techniques, enabling the capture of distinct behaviours within the data. The broken power law, given by two equations for different ranges of x, introduces more complexity. It is a piecewise function given by a sequence of conjoined power laws (x) where each section has its own power (b) and is defined by bounding "breaks" .
For $x \leq x_b$, the relationship is $S_x = a_1 x^{b_1}$, while for $x > x_b$, it changes to $S_x = a_2 x^{(b_1-b_2)/b} x^{b_2}$. Here, $x_b$ represents the breakpoint where the behaviour of the relationship changes. Broken power law fitting allows for a different set of parameters $a_1$, $b_1$, $a_2$, $b_2$ on either side of this breakpoint, providing a much more flexible model to capture two different behaviours in the data.

**Data Analysis**

Open source python programs were also frequently used to organise the large amount of data and to plot the graphs, as well as to conduct the broken power law analysis.

**REFERENCES**

1. R. C. Smith, "Cataclysmic variables," arXiv.org, Jan. 23, 2007. https://arxiv.org/abs/astro-ph/0701654
2. S. N. Shore, "Binary Stars," in Elsevier eBooks, 2003, pp. 77–92. doi: 10.1016/b0-12-227410-5/00052-1.
3. R. D. Gehrz, G. L. Grasdalen, J. A. Hackwell, and E. P. Ney, "The evolution of the dust shell of Nova Serpentis 1978," The Astrophysical Journal, vol. 237, p. 855, May 1980, doi: 10.1086/157934.
4. M. Panda and Y. Chandra, "Unveiling the Mysteries of the Cosmos: An overview of radio astronomy and its profound insights," arXiv (Cornell University), Aug. 2023, doi: 10.48550/arxiv.2308.09415.
5. L. Chomiuk *et al.*, "Classical Novae at Radio Wavelengths," *Astrophysical Journal Supplement Series*, vol. 257, no. 2, p. 49, Dec. 2021. doi: 10.3847/1538-4365/ac24ab.
6. "Very large array - National Radio Astronomy Observatory," National Radio Astronomy Observatory, Mar. 24, 2023. https://public.nrao.edu/telescopes/vla/
7. "The Karl G. Jansky Very Large Array — Science website." https://science.nrao.edu/facilities/vla
8. A. Mittal, A. Santra, V. Bhatnagar, and D. Khanna, "Exploratory Analysis of Light Curves: A Case-Study in Astronomy Data Understanding," in Lecture Notes in Computer Science, 2014, pp. 67–94. doi: 10.1007/978-3-319-05693-7_5.
9. M. F. Bode and A. Evans, *Classical Novae*. 1989.
10. K. R. Lang, Astrophysical Formulae: Space, Time, Matter and Cosmology. Springer, 2013.
11. K. McDonnell, M. Wardle, and A. E. Vaughan, "A search for OH 6 GHz maser emission towards supernova remnants," *Monthly Notices of the Royal Astronomical Society*, Oct. 2008. doi: 10.1111/j.1365-2966.2008.13728.x.
12. J. Zhao, K. R. Anantharamaiah, W. M. Goss, and F. Viallefond, "Radio Recombination Lines from the Nuclear Regions of Starburst Galaxies," *The Astrophysical Journal*, vol. 472, no. 1, pp. 54–72, Nov. 1996. doi: 10.1086/178041.
13. B. C. Lacki, "Interpreting the low-frequency radio spectra of starburst galaxies: a pudding of Strömgren spheres," *Monthly Notices of the Royal Astronomical Society*, vol. 431, no. 4, pp. 3003–3024, Apr. 2013. doi: 10.1093/mnras/stt349.
14. J. J. Condon and S. M. Ransom, *Essential radio astronomy*. Princeton University Press, 2016.

15. L. Takeda *et al.*, "Optical and near-infrared data and modelling of nova V5668 Sgr," *Monthly Notices of the Royal Astronomical Society*, vol. 511, no. 2, pp. 1591–1600, Jan. 2022. doi: 10.1093/mnras/stac097.
16. D. P. K. Banerjee, M. Srivastava, N. M. Ashok, and V. Venkataraman, "Near Infrared studies of the carbon-monoxide and dust forming nova V5668 Sgr," *arXiv (Cornell University)*, Oct. 2015. doi: 10.48550/arxiv.1510.04539.
17. D. Jack *et al.*, "Study of the variability of Nova V5668 Sgr, based on high-resolution spectroscopic monitoring," *Astronomische Nachrichten*, vol. 338, no. 1, pp. 91–102, Jan. 2017. doi: 10.1002/asna.201613217.
18. L. Chomiuk, B. D. Metzger, and K. J. Shen, "New Insights into Classical Novae," *Annual Review of Astronomy and Astrophysics*, vol. 59, no. 1, pp. 391–444, Sep. 2021. doi: 10.1146/annurev-astro-112420-114502.
19. R. D. Gehrz, J. W. Truran, R. E. Williams, and S. Starrfield, "Nucleosynthesis in classical Novae and its contribution to the interstellar medium," Publications of the Astronomical Society of the Pacific, vol. 110, no. 743, pp. 3–26, Jan. 1998. doi: 10.1086/316107.
20. M. Friedjung, "Mass Loss from Novae and Supernovae," Highlights of Astronomy, vol. 6, pp. 581–592, Jan. 1983. doi: 10.1017/s1539299600005591.
21. E. Santamaría et al., "Angular expansion of nova shells," The Astrophysical Journal, vol. 892, no. 1, p. 60, Mar. 2020. doi: 10.3847/1538-4357/ab76c5.
22. M. Maggini, "The change in brightness, spectrum, and temperature of Nova Aquilae no. 3.," The Astrophysical Journal, vol. 48, p. 303, Dec. 1918. doi: 10.1086/142437.

**FIGURES, TABLES, AND CAPTIONS**

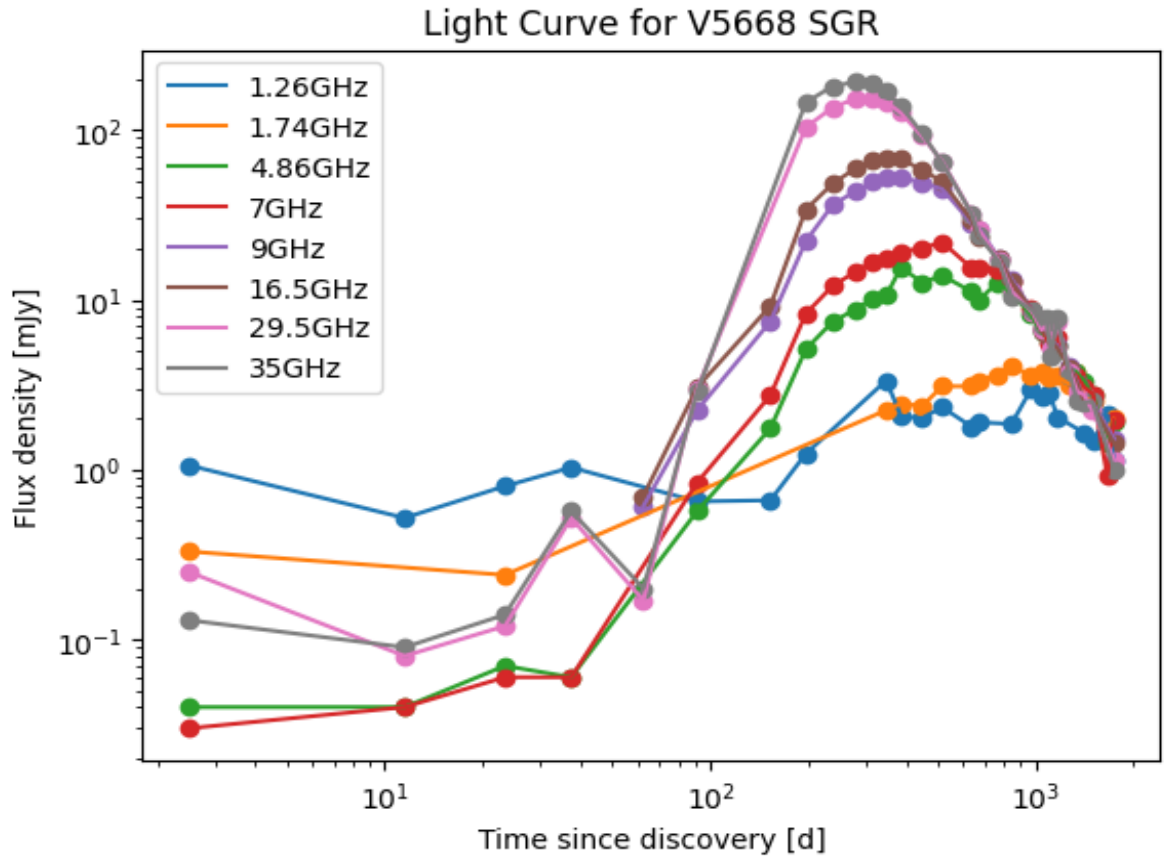

*Figure 1.* **Graph Of The Light Curves. The function of flux density over time of the range of frequencies of Nova V5668 SGR.**

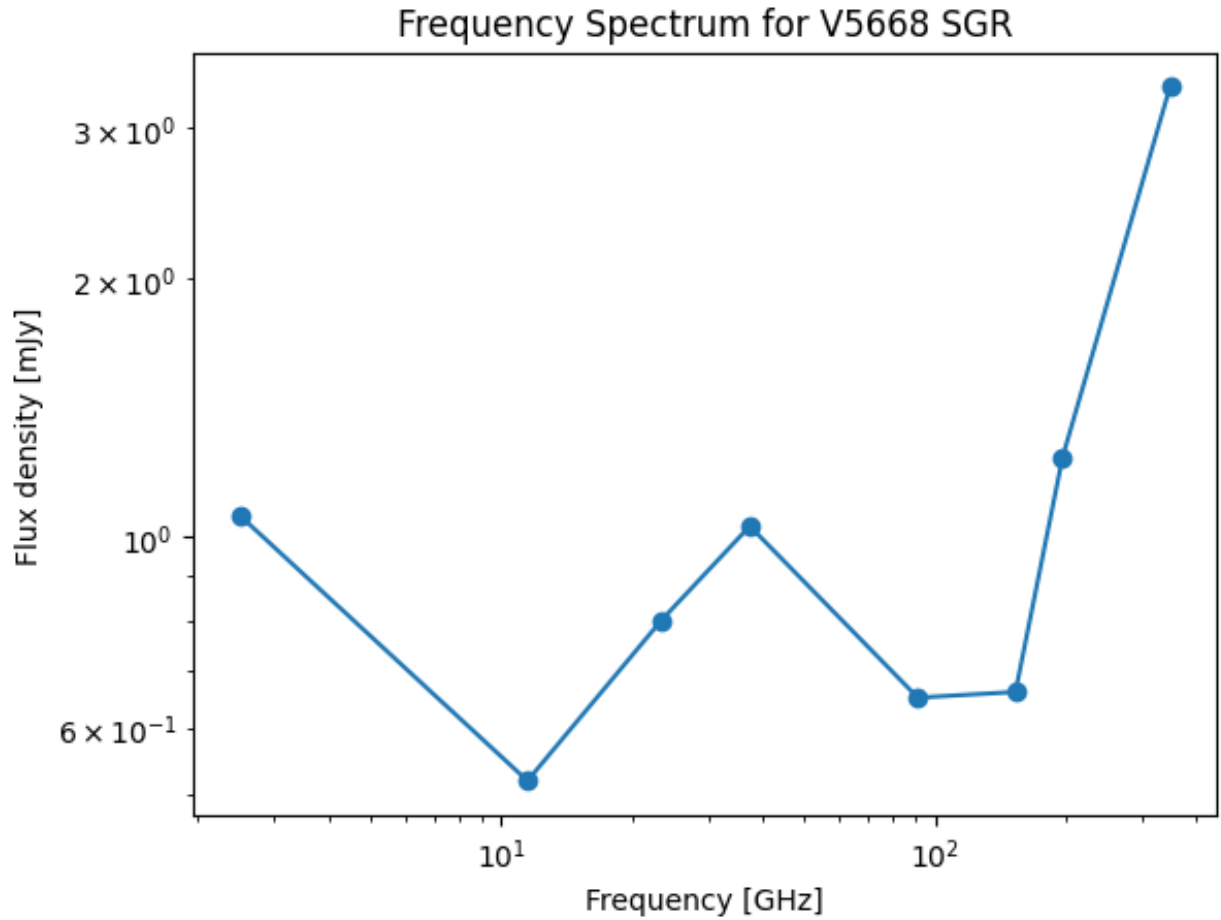

*Figure 2.* **Graph of the Frequency Spectrum. It shows the frequency spectrum of Nova V5668 SGR for the Day of observation 635.8. Frequencies ranged from 1.26 GHz to 35 GHz.**

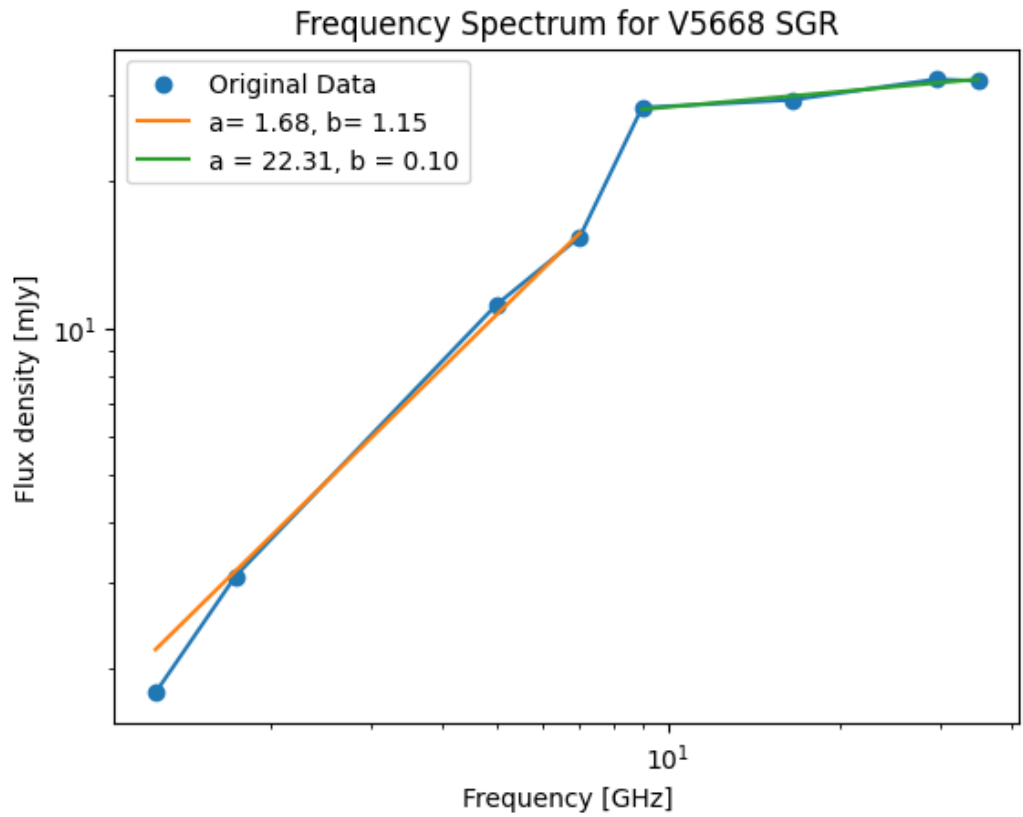

*Figure 3.* **Broken Power Law Fitted Graph. It shows the frequency spectrum plot of nova V5668 Sgr fitted with broken power law. The a and b values for the broken power law fitting are displayed in the figure.**